\documentclass[aps,pre,superscriptaddress,twocolumn,longbibliography]{revtex4}

\usepackage{graphicx,epstopdf,url}
\usepackage[colorlinks=true,urlcolor=blue,citecolor=blue,linkcolor=blue,urlcolor=blue]{hyperref}

\usepackage[active]{srcltx}

\begin{document}
\title{
Chiral organic molecular structures supported by multilayer surfaces
}

\author{Alexander V. Savin}
\email[]{asavin@chph.ras.ru}
\affiliation{Semenov Institute of Chemical Physics, Russian Academy of Sciences,
Moscow 119991, Russia}
\affiliation{
Plekhanov Russian University of Economics, Moscow 117997, Russia
}

\author{Yuri S. Kivshar}
\email[]{yuri.kivshar@anu.edu.au}
\affiliation{Nonlinear Physics Center, Department of Fundamental and Theoretical Physics, Research School of Physics, Australian National University, Canberra ACT 2601, Australia}


\begin{abstract}
We study numerically the dynamics of acetanilide  (ACN) molecules placed on a flat surface of a multilayer hexagonal boron nitride  structure. We demonstrate that the ACN molecules,  being achiral in three dimensions, become chiral after being placed on the substrate. Homochirality of the ACN molecules leads to stable secondary structures stabilized by hydrogen bonds between peptide groups of the molecules. Numerical simulations of systems of such molecules reveal that the structure of the resulting hydrogen-bond chains depends on the isomeric composition of the molecules.
If all molecules are homochiral (i.e. only one isomer is present), they form secondary structures  (chains of hydrogen bonds in the shapes of arcs, circles, and spirals). If the molecules at the substrate form a racemic mixture, then no regular secondary structures appear, and only curvilinear chains of hydrogen bonds of random shapes can emerge. A hydrogen-bond chain can form a straight zigzag only if it has an alternation of isomers. Such chains can create two-dimensional (2D) regular lattices, or 2D crystals. The melting scenarios of such 2D crystals depend on density of its coverage of the substrate.  At 25\% coverage, melting occurs continuously in a certain temperature interval. For a complete coverage, melting occurs  at $415\div 470$~K due to a shift of 11\% of all molecules into the second layer of the substrate.
\end{abstract}

\maketitle

\section{Introduction}

Two-dimensional (2D) materials such as graphene (G) and hexagonal boron nitride (h-BN)  have attracted a lot of attention due to their unique electronic~\cite{Novoselov2004,Neto2009,Koren2016} and mechanical~\cite{Meyer2007,Lee2008,Falin2017,Han2020} properties.
Currently, heterogeneous layered materials of such 2D materials, which can exhibit various novel physical properties compared to their homogeneous counterparts, became a special focus of such studies~\cite{Leven2013,Geim2013,Novoselov2016}.
For example, the use of hybrid G/h-BN structures allows to achieve some desired electronic properties~\cite{Woods2014,Slotman2015} and also reduce significantly friction between the layers~\cite{Mandelli2017}. In general, such multilayer heterostructures are stabilized by van der Waals (vdW) interactions between atoms of the neighboring layers.

The concept of vdW heterostructures can be extended to the integration of 2D materials with molecular structures of different dimensions, such as $n$D/2D heterostructures,  where $n$ stands for the dimension ($n=0$, 1, or 3)~\cite{Jariwala2017}, describing flat molecules ($n=0$), polymer chains ($n=1$), or three-dimensional molecular objects ($n=3$).

\begin{figure}[tb]
\begin{center}
\includegraphics[angle=0, width=1.0\linewidth]{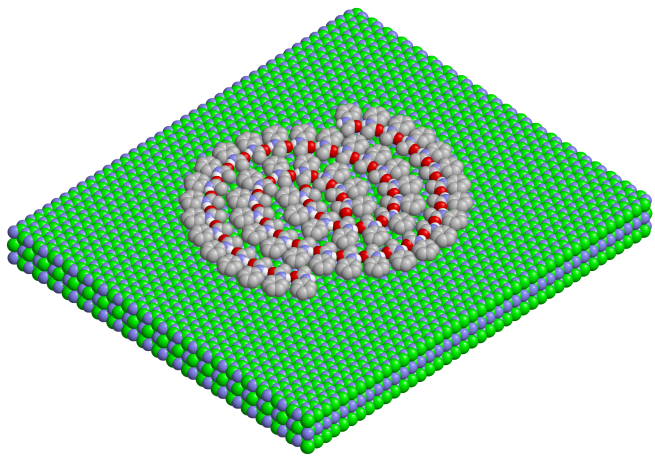}
\end{center}
\caption{\label{fg01}\protect
Example of the molecular structures studied in this paper.  A spiral structure of 52 R-isomers of the ACN molecules C$_6$H$_5$--NHCO--CH$_3$ is stabilized on a h-BN multilayer surface.
}
\end{figure}

For molecules and molecular chains with benzol rings, flat layers of G and h-BN are strong adsorbents~\cite{Ershova2010,Gordeev2013,Wang2014,Castro2014,Zhou2015}.
Theoretical studies reveal that molecules adsorbed on G and h-BN surfaces through non-covalent  interactions can modify the properties of the surface as solid-liquid, solid-air,  or solid-vacuum interfaces~\cite{Georgakilas2012,Georgakilas2016,Thakkar2022}.
A strong stacking interaction with a flat substrate allows such molecules residing on a surface creating stable 2D supra-molecular systems, as shown in the characteristic example of Fig.~\ref{fg01}

By now, the behavior of multifunctional organic molecules placed on ideal metal surfaces has been studied in detail~\cite{Barlow2003}. Such organometallic systems may exhibit a variety of different structures induced by the substrate. In many cases, complex organic molecules (such as carboxylic acids, amino acids, anhydrides, and ring systems)  become self-organized on metal surfaces creating ordered super-structures stabilized by inter-molecular interactions. Chirality is of a particular interest that can appear for initially achiral metal surfaces by adsorbing organic molecules~\cite{Cao2016}. Similar behavior is expected for organic molecules adsorbed on flat surfaces of  G and h-BN molecular structures. In this paper, we study numerically the formation of supra-molecular complexes by acetanilide molecules placed on the surface of a multilayer h-BN sheet, see Fig.~\ref{fg01}, serving as an introductory figure explaining our problem and results discussed below.

For poly-cyclic aromatic hydrocarbons (for molecules of benzol C$_6$H$_6$, naphthalene C$_{10}$H$_8$, pyrene C$_{16}$H$_{10}$,...), graphene is a strong adsorbent \cite{Ershova2010,Wang2014,Zhou2015,Bahn2014}. The interaction of graphene with such molecules often causes specific reactions that can be used  in new types of sensors \cite{Schedin2007,Shao2010}. Non-covalent functionalization of the graphene surface can significantly expand its potential  range of applications \cite{Georgakilas2012,Georgakilas2016}. It has been shown experimentally~\cite{Bahn2014,Zhen2018} that benzol and pyrene molecules
adsorbed on graphene form densely packed monolayers.

For acetanilide (ACN, C$_6$H$_5$NHCOCH$_3$) and paracetamol (PCM, C$_6$H$_4$OHNHCOCH$_3$)  molecules, graphene and hexagonal boron nitride are also strong adsorbents. Due to possible medical applications, much attention has been paid to modeling the adsorption  of PCM molecules, which is a strong analgesic, on h-BN sheets and nanotubes \cite{Castro2014,Iranmanesh2019}. It has been shown in \cite{Kang2010} that functionalized graphene can be used
as a highly sensitive paracetamol detection sensor.

An example of a 1D/2D heterostructure is a graphene sheet with adsorbed Kevlar chains,  kevlar-functionalized graphene  \cite{Lian2014}. The presence of planar C$_6$H$_4$ benzol rings and NHCO peptide groups in the polymer chain
[--C$_6$H$_4$--NHCO--]$_\infty$ provides a strong non-covalent (vdW) interaction of the chain  with G and h-BN sheets. Such chains on the surface of sheets G and h-BN will lie parallel to the surface and form  chains of hydrogen bonds between each other $\cdots$HNCO$\cdots$HNCO$\cdots$.

Such 3D/2D heterostructures can form thin metal layers on the G and h-BN surfaces. In particular, numerical modeling suggests that aluminum can form stable two-layer structures on the G surface~\cite{Kumar2017}.

It has been shown in Refs.~\cite{Roth2005,Pint2006,Becker2006,Chen2008,Connolly2008,Yang2008,Roth2016,Piskorz2019,Fang2019,Zhang2021} that n-alkanes (linear polymer chains CH$_3$(CH$_2$)$_l$CH$_3$ with internal units $2\le l\le 388$) form a dense ordered monolayer of parallel linear chains on the graphite (graphene) surface.
Interest in alkanes is due to the fact that they belong to the simplest families of polymer  molecules, which members of which differ only in their length.
Placing linear polymer chains on a flat graphite surface causes them to self-assemble into 2D crystals. The self-assembly mechanism depends on the chain length, temperature,  and the level of coverage of the substrate with chains \cite{Chen2008,Piskorz2019}.

Adsorption by the surface of a long single-chain polyethylene molecule leads to its two-dimensional crystallization -- it passes from the form of a three-dimensional globule into the form of  a parallel folded linear chain lying in the plane parallel to the substrate surface \cite{Yang2011,Gulde2016,Liu2018}.

Thus, the flat surfaces of the G and h-BN substrates create a 2D platform for flat molecules adsorbed  on them (for poly-cyclic aromatic hydrocarbons, for ACN and PCM molecules, Kevlar chains, ...)  and linear polymer molecules. At low temperatures, the molecules move along the sheet, remaining parallel to its surface. They interact with each other and form two-dimensional supra-molecular structures. Such molecular adsorbents are convenient systems for studying phase transitions caused by freedom restrictions.

To date, only phase transitions in monolayers of n-alkanes have been well studied
\cite{Roth2005,Pint2006a,Yang2006,Wexler2009,Zhang2021}.
Modeling and experimental studies show that a monolayer always undergoes a transition from  a solid-crystalline 2D phase to a liquid phase (the transition occurs at a temperature  significantly lower than the desorption temperature of molecules). The melting scenario depends on the polymer chain length. The melting temperature increases  monotonically with chain length, so for pentane, heptane and nonane ($l=3$, 5, 7) the melting  temperature is $T_m=92$, 178 and 255K  \cite{Pint2006a}. A characteristic feature associated with the adsorption of molecules is the continuity  of melting of a 2D crystal -- melting occurs in the temperature interval. Thus, for the longest synthesized monodisperse alkane C$_{390}$H$_{782}$ ($l=388$), continuous melting occurs at $393<T<484$~K \cite{Zhang2021}.

Despite the large number of theoretical and experimental works on phase transitions
in adsorbed monolayers of alkanes and their derivatives, as far as we know,
there are no works on modeling phase transitions in adsorbed monolayers of ACN, PCM,
and Kevlar (para-aramid) molecules. A detailed description of adsorption simulation methods is given in \cite{Pykal2016}. Unlike alkanes, the 2D structures of these molecules adsorbed on a flat surface  are associated with the presence of chains of hydrogen bonds. Molecules including amide and hydroxyl groups can create 2D lattices and extended hydrogen chains.

We notice that the ACN molecules are often considered as a model system with chains of hydrogen  bonds between HNCO peptide groups. Acetanilide crystallizes into an orthorhombic structure with ribbons of molecules  linked by hydrogen bonds \cite{Johnson1995}. The chains of hydrogen bonds that stabilize  the crystal structure are very similar to the chains that stabilize the alpha-helices  and beta-sheets of proteins. Therefore, ACN was used as a model for modeling the energy transfer of vibrations  of peptide groups along hydrogen bond chains in proteins \cite{Careri1984,Scott1992,Cruzeiro2013}.
\begin{figure}[tb]
\begin{center}
\includegraphics[angle=0, width=0.6\linewidth]{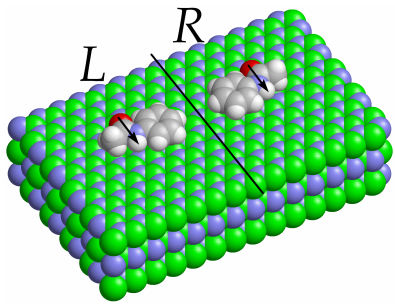}
\end{center}
\caption{\label{fg02}\protect
When an ACN molecule is placed on a flat surface of a multilayer
h-BN structure, it may create two isomers with the mirror symmetry (shown by a straight line). Vectors connecting the oxygen atom with the hydrogen atom of each peptide group show the dipole moments. For the L-isomer, the benzol ring is to the left of this vector, for the R-isomer it is to the right. Gray balls stand for carbon atoms, white balls -- hydrogen, blue -- nitrogen, red -- oxygen, and green -- bromine atoms.
}
\end{figure}

Living matter, unlike non-living matter, has chiral purity: all proteins consist of left-handed  amino acids, while DNA and RNA are built on right-handed ribose.
In experiments on abiogenic synthesis, left and right isomers of sugars and proteins
are formed in equal proportions. It is believed that if you try to build proteins from such a mixture,  they will not be able to fold into a stable form and therefore will not work as enzymes. In three dimensions, the need for chiral purity to form stable protein structures requires complex analysis. The situation is dramatically simplified if we move from three-dimensional space to  two-dimensional. Such a transition can be made if flat molecules are placed on a flat molecular
sheet of graphene or hexagonal boron nitride (h-BN). Such a nonvalent modification of the sheet surface actually creates a 2D world  for the flat molecules placed on it. At low temperatures, the molecules move along the sheet all the time remaining parallel to its surface. On the surface, they can form complex two-dimensional structures.

An ACN molecule that is achiral in 3D becomes chiral after being placed on a flat substrate  (the chirality depends on which side it lays on the surface of the sheet) -- see Fig.~\ref{fg02}. It will be shown that the homochirality of ACN molecules leads to the appearance on the surface  of the sheet of stable secondary structures stabilized by hydrogen bonds: cyclic  and spiral chains and complexes of them. Modeling the formation of such structures will make it possible to demonstrate the necessity  of homochirality (chiral purity) of biomolecules for the formation of stable secondary  molecular structures from them.

As a flat substrate, we consider a surface of a multilayer h-BN structure, and for molecules we consider acetanilide (ACN) C$_6$H$_5$NHCOCH$_3$, as shown in Figs.~\ref{fg01} and \ref{fg02}. The presence of a planar benzol ring C$_6$H$_5$ and a planar peptide group (PG) HNCO leads to large interaction energy of the molecule with the substrate, $E_{sub}=0.762$~eV. Molecules can create chains of hydrogen bonds between their peptide groups OCNH$\cdots$OCNH$\cdots$OCNH$\cdots$. Such chains of hydrogen bonds stabilize the secondary structures of the protein molecules.

The paper is organized as follows. In the next section, we describe our model.
Section III is devoted to the study of secondary structures of the ACN molecules placed on a flat substrate. Self-assembly of such structures is simulated numerically in Sec. IV. Then, in Sec. V we analyze melting of 2D crystals. Section VI concludes the paper.

\section{Model}

\begin{figure}[tb]
\begin{center}
\includegraphics[angle=0, width=1.\linewidth]{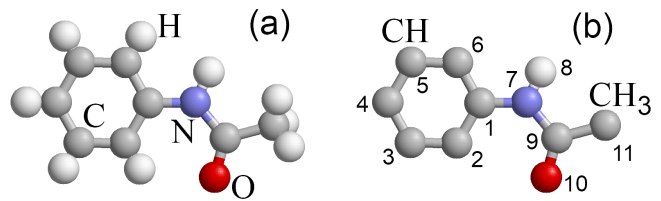}
\end{center}
\caption{\label{fg03}\protect
Construction of a coarse-grained model of the ACN molecule:
(a) full-atomic view of the molecule, (b) coarse-grained model
(the used numbering of the united atoms is shown).
}
\end{figure}

For modeling of the dynamics of a system of the ACN molecules, we will use the united-atoms approximation. Let us consider the molecular groups CH and CH$_3$ as united atoms whose centers coincide  with the centers of carbon atoms. In this approximation, the ACN molecule is described as a system of 11 united atoms -- see Fig.~\ref{fg03}.
The values of the masses of the united atoms are shown in the table \ref{tab1}.

To model a ACN molecule, we use the force field in which distinct potentials describe the deformation of valence bonds and valence, torsion and dihedral angles, and non-valence atomic interacts \cite{Cornell1995}. In this model, the deformation energy of the valence bonds
C--CH, CH--CH, C--N, N--H, C=O and C--CH$_3$  is described by the harmonic
potential:
\begin{equation}
V(\rho)=\frac12K(\rho-\rho_0)^2,
\label{f1}
\end{equation}
where $\rho$ and $\rho_0$ are current and equilibrium bond lengths, $K$ is the bond stiffness.
The values of potential parameters for various valence bonds are presented in Table \ref{tab2}.
\begin{table}[b]
\caption{
Masses and parameters of interaction potentials for united atoms X of the ACN molecule:
$i$ -- atom number, $M_i$ -- atom mass ($m_p=1.6603\times 10^{-27}$~kg -- proton mass),
$\varepsilon_i$ and $r_i$ are the energy and radius of the LJ interaction,
$q_i$ is the electric charge of the atom,
$\epsilon_i$ and $h_i$ are the energy and equilibrium distance for the interaction
of an atom with a flat substrate (with a crystal surface h- BN).
\label{tab1}
}
\begin{center}
\begin{tabular}{c|ccccccc}
 ~X~                  & C    & CH        &  N    & H     & C     & O     & CH$_3$\\
 $i$                  & 1    & 2,3,4,5,6 &  7    & 8     & 9     & 10    & 11\\
 \hline
~$M_i$~($m_p$)        & 12   & 13       &  14    & 1     & 12    & 16    & 15 \\
$\varepsilon_i$~(meV) & ~4.284 & 4.284   & 4.080  & 0.434 & 4.284 & 6.344 & 4.284 \\
$r_i$~(\AA)           & ~1.861 & 1.861   & 1.899  & 0.621 & 1.861 & 1.711 & 1.861 \\
$q_i$~($e$)           & ~0.066 & 0       & -0.463 & 0.286 & 0.580 & -0.504 & 0.035 \\
$\epsilon_i$~(meV)    & ~61.5  & 87.3    & 47.7   & 31.3  & 61.5  & 42.8   & 87.3 \\
$ h_i$~(\AA)          & ~3.52  & 3.44    & 3.43   & 3.08  & 3.52  & 3.36   & 3.44\\
\hline
\end{tabular}
\end{center}
\end{table}
\begin{table}[b]
\caption{Values of the harmonic potential parameters (\ref{f1}) for different
valence bonds X---Y. \label{tab2}
}
\begin{center}
\begin{tabular}{l|ccccc}
 ~X---Y~    & C--CH,~CH--CH &  C--N & N--H  & C=O  & C--CH$_3$\\
\hline
~$K$~(N/m)~ & 469           & 427   & 434   & 570   & 317 \\
~$\rho_0$~(\AA)& 1.39       & ~1.405~ & ~1.007~ & ~1.222~ & ~1.505~\\
\hline
\end{tabular}
\end{center}
\end{table}
\begin{table*}[t]
\caption{Values of the parameters of the potential of the valence angle X--Y--Z (\ref{f2}) for different
atoms. \label{tab3}
}
\begin{center}
\begin{tabular}{l|ccccccc}
 ~X--Y--Z~              & ~C--C--C~ & ~C--C--N~ & ~C--N--H~ & ~C--N--C~ & ~N--C--O~ & ~N--C--C~ & ~O--C--C~\\
 \hline
 $\epsilon_{a}$~(eV) &  3.643     & 3.823     & 2.781    & 4.888     & 4.932 & 3.758 &  4.625\\
 $\varphi_0$~($^\circ$)& 120      & 117       & 118      & 128       & 123  &  116 &  120 \\
\hline
\end{tabular}
\end{center}
\end{table*}
\begin{table*}[t]
\caption{Values of the parameters of the potential of the dihedral angle X--Y--Z--W (\ref{f3}) for different
atoms. \label{tab4}
}
\begin{center}
\begin{tabular}{l|cccccc}
~X--Y--Z--W~          & ~C--C--C--C~ & ~C--C--C--N~ & ~C--C--N--H~ & ~C--C--N--C~ & ~C--N--C--O~ & ~C--N--C--CH$_3$~\\
 \hline
 $\epsilon_{d}$~(eV) & 0.63       & 0.63       & 0.21       & 0.21       & 0.42       &  0.42\\
 $z_d$               &  -1        &  1         & -1         & -1         & -1 &  1 \\
 $k$                 &   1        &  1         &  2         &  2         &  1 &  1 \\
\hline
\end{tabular}
\end{center}
\end{table*}

Energies of the deformation of the valence angles X--Y--Z are described by the potential
\begin{equation}
U({\bf u}_1,{\bf u}_2,{\bf u}_3)=U(\phi)=\epsilon_a(\cos\phi-\cos\phi_0)^2,
\label{f2}
\end{equation}
where the cosine of the valence angle $\phi$ is defined as
$\cos\phi=-({\bf v}_1,{\bf v}_2)/\rho_1\rho_2$, with the vectors
${\bf v}_1={\bf u}_2-{\bf u}_1$, ${\bf v}_2={\bf u}_3-{\bf u}_2$ and bond lengths
$\rho_1=|{\bf v}_1|$, $\rho_2=|{\bf v}_2|$, the vectors ${\bf u}_1$, ${\bf u}_2$, ${\bf u}_3$
specify the coordinates of the atoms forming the valence angle $\phi$, $\phi_0$ is
the value of equilibrium valence angle. The values of potential parameters used for various
valence angles are presented in Table~\ref{tab3}.

Deformation of dihedral angles are described by the potential
\begin{equation}
W({\bf u}_1,{\bf u}_2,{\bf u}_3,{\bf u}_4)=\epsilon_d(1+z_d\cos k\varphi),
\label{f3}
\end{equation}
where $\cos\varphi=({\bf w}_1{\bf w}_2)/|{\bf w}_1||{\bf w}_2|$,
with the vectors ${\bf w}_1=({\bf u}_2-{\bf u_1})\times ({\bf u}_3-{\bf u_2})$,
${\bf w}_2=({\bf u}_3-{\bf u_2})\times ({\bf u}_4-{\bf u_3})$.
The values of potential parameters used for various dihedral angles are presented in Table~\ref{tab4}.
\begin{figure}[tb]
\begin{center}
\includegraphics[angle=0, width=0.6\linewidth]{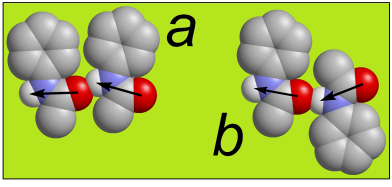}
\end{center}
\caption{\label{fg04}\protect
Dimer of (a) RR and (b)LR isomers of the ACN molecule on a flat substrate (shown in green).
The vectors show the dipole moments of the peptide groups.
Hydrogen bond energy of identical isomers $E_{hb}=0.330$~eV,
angle between dipole moments $\phi_{hb}=17.05^\circ$,
for different isomers $E_{hb}=0.322$~eV, $\phi_ {hb}=33.98^\circ$.
}
\end{figure}

For pairs of atoms X$_i$,$X_j$ ($i,j$ are the numbers of atoms in the molecule) participating
in the formation of the dihedral angle X$_i$--Y--Z--X$_j$, their non-valence interaction
is also taken into account described by the Lennard-Jones (LJ) potential
\begin{equation}
W_0(r)=\varepsilon_0[(r_0/r)^{12}-2(r_0/r)^6],
\label{f4}
\end{equation}
with halved interaction energy $\varepsilon_0=\sqrt{\varepsilon_i\varepsilon_j}/2$,
where $r$ is current distance between interacting atoms, equilibrium distance $r_0=r_i+r_j$.
The LJ interaction of an oxygen atom ($i=10$) with two combined atoms CH ($i=2,6$) was also
taken into account with interaction energy $\epsilon_0=\sqrt{\varepsilon_2\varepsilon_{10}}$
and equilibrium distance $r_0=r_2+r_{10}$. Parameter values $\varepsilon_i$ and $r_i$
are shown in Table~\ref{tab1}.

The interaction of two ACN molecules is described by the potential
\begin{eqnarray}
U({\bf X}_1,{\bf X_2})=\sum_{i=1}^{11}\sum_{j=1}^{11}\{\varepsilon_{ij}[(\bar{r}_{ij}/r_{ij})^{12}-2(\bar{r}_{ij}/r_{ij})^6]
\nonumber\\
+\kappa q_iq_j/r_{ij}\}, \label{f5}
\end{eqnarray}
where the 33-dimensional vector ${\bf X}_k=\{ {\bf u}_{k,i}\}_{i=1}^{11}$ ($k=1,2$)
defines the coordinates of atoms of the $k$-th ACN (vector ${\bf u}_{k,i}$ specifies the
coordinates of the $i$-th atom of the $k$-th molecule),
distance between atoms $r_{ij}=|{\bf u}_{1,i}-{\bf u}_{2,j}|$.
Here energy $\varepsilon_{ij}=\sqrt{\varepsilon_i\varepsilon_j}$,
equilibrium distances $\bar{r}_{ij}=r_i+r_j$, $q_i$ is the electric charge of $i$-th atom
($i,j=1,...,11$), coefficient $\kappa=14.400611$~eV\AA/$e^2$.
The values of the parameters $\varepsilon_i$, $r_i$ and $q_i$ are shown in Table~\ref{tab1}.
All values of parameters of interaction potentials (\ref{f1}), (\ref{f2}), (\ref{f3}) and (\ref{f5})
are obtained from force field AMBER \cite{Cornell1995}.

The van der Waals interactions of the atoms of the ACN molecule with flat substrate are described
by the LJ potential $(m,l)$
\begin{equation}
W({\bf X})=\sum_{i=1}^{11}W_i(z_i)=\sum_{i=1}^{11}\frac{\epsilon_i}{l-m}[m(h_i/z_i)^l-n(h_i/z_i)^m],
\label{f6}
\end{equation}
where $z_i$ is the distance from $i$-th atom to the outer surface of the  substrate,
which is plane $z=0$.
Potential $W_i(z_i)$ in Eq. (\ref{f6}) is the interaction energy of $i$-th atom as a function
of the distance to the substrate. This energy was found numerically for different substrates
\cite{Savin2019,Savin2021}. The calculations showed that  interaction energy with substrate
$W_i(z)$ can be described with a high accuracy by LJ potential (\ref{f6}) with power $l>k$.
Potential $W_i(z)$ has the minimum $W_i(h_i)=-\epsilon_i$ ($\epsilon_i$ is the binding energy
of the $i$-th atom with substrate). For the surface of the h-BN crystal $l=10$, $m=4.25$.
The values of the parameters $\epsilon_i$, $h_i$, $i=1,...,11$, are given in the table \ref{tab1}.
The 10-layer fragment of h-BN crystal was used to find  values of this parameters.
The interaction energy of an atom with a substrate was found as the sum of all LJ potentials
(\ref{f4}) with parameters from the force field UFF \cite{Rappe1992}.

Thus, the Hamiltonian of a system of $N$ ACN molecules located on the flat surface of
h-BN crystal has the form
\begin{equation}
H=\sum_{n=1}^N\frac12({\bf M}\dot{\bf X}_n,\dot{\bf X}_n)+P,\label{f7}
\end{equation}
where the first term specifies the kinetic and the second -- potential energy of the system
\begin{equation}
P=\sum_{n=1}^N[V({\bf X}_n)+W({\bf X}_n)]
+\sum_{n=1}^{N-1}\sum_{k=n+1}^N U({\bf X}_n,{\bf X}_k).
\label{f8}
\end{equation}
Here the vector ${\bf X}_n=\{ {\bf u}_{n,i}\}_{i=1}^{11}$ specifies the coordinates of the atoms
of $n$-th ACN molecule,
${\bf M}$ is the diagonal matrix of atom masses of the molecule,
$V({\bf X}_n)$ and $W({\bf X}_n)$ are deformation energy and energy of interaction with the substrate
of $n$-th molecule,
$U({\bf X}_n,{\bf X}_k)$ is the interaction energy of $n$ and $k$ molecules.

\section{Secondary structures of ACN molecules on a flat substrate}

The ACN molecule is achiral, but it becomes chiral when placed on a flat substrate.
Depending on which side it lies on the substrate, it can be either right
(when the benzol ring is located to the right of the dipole moment vector of the peptide group
$\vec{\rm OH}$) or left (the benzol ring is located to the left).
Two mirror-symmetrical isomers of the molecule are shown in Fig.~\ref{fg02}.
To transfer a molecule from one isomer to another, it must be partially torn off the substrate
and be placed on the substrate with its other side.
All this requires overcoming the energy barrier $\Delta E=0.466$~eV.
The total energy of interaction with the h-BN substrate (desorption energy) $E_{sub}=0.762$~eV.
Therefore, the spontaneous transition of the ACN molecule lying on the substrate from the L
to the R form and vice versa is possible only at temperatures $T>300$~K.
At lower temperatures, the molecule will always stay on the substrate,
adjoining it with the same side, i.e. without changing the isomer type.

To find the stationary state of the system of  $N$ ACN molecules lying on a flat h-BN substrate,
it is necessary to find the state of the system with a minimum potential energy
\begin{equation}
P\rightarrow \min: \{ {\bf X}_n\}_{n=1}^N.
\label{f9}
\end{equation}
The minimization problem (\ref{f9}) is solved numerically by the conjugate gradient method.
Choosing the starting point of the minimization procedure, one can obtain all the
main stationary states of the molecular system.
\begin{figure}[tb]
\begin{center}
\includegraphics[angle=0, width=1.0\linewidth]{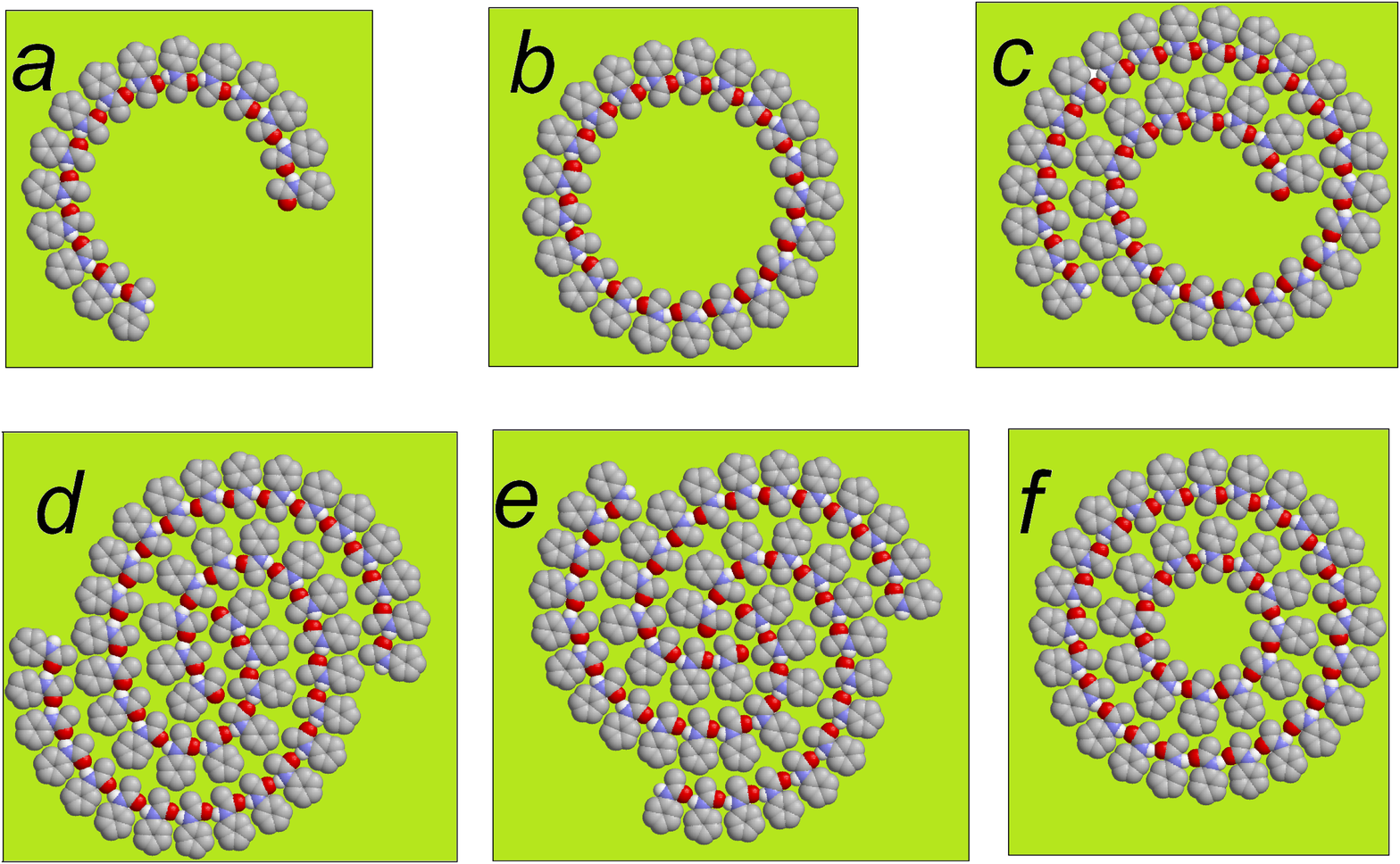}
\end{center}
\caption{\label{fg05}\protect
Typical secondary structures of homochiral ACN molecules on a flat substrate:
(a) arc (number of atoms $N=15$), (b) circle ($N=21$),
(c) single-beam spiral ($N=34$), (d) two-beam spiral ($N=23+23$),
(e) three-beam spiral  ($N=15+15+15$), (f) nested circles ($N=12+24$).
}
\end{figure}

Peptide groups of neighboring molecules can form hydrogen bonds, creating dimers -- see Fig.~\ref{fg04}.
The numerical solution of the problem (\ref{f9}) shows that when molecules are located on a flat substrate,
two types of dimers are possible: dimers of molecules of the same and different chirality.
If a dimer is formed by identical isomers, then its binding energy is slightly higher:
the hydrogen bond energy for RR and LL isomers is $E_{hb}=0.330$~eV, and for RL and LR isomers $E_{hb}=0.322$~eV.
This is due to the fact that in this case the benzol rings C$_6$H$_5$ of the molecules
are on the same side and they make a larger contribution to the interaction energy.
The hydrogen bond angle also depends on the chirality of the dimer molecules.
For dimer molecules of the same chirality, the angle between the dipole moments of the peptide
groups forming a hydrogen bond is $\phi_{hb}=17^\circ$, and for molecules of different
chirality $\phi_{hb}=34^\circ$.

Chains of hydrogen bonds of molecules of the same chirality will always have benzol rings
on one (outer) side, so they will twist in the opposite (inner) direction and have
approximately the same curvature.
On a plane, circular arcs, circles, and spirals have such properties.
The solution of the problem (\ref{f9}) has shown that on a flat substrate molecules
of the same chirality form stable shape structures with little changing curvature:
arcs, spirals, circles -- see Fig.~\ref{fg01} and \ref{fg05}.
Left isomers form structures with a twist to the right, right -- to the left.
\begin{figure}[tb]
\begin{center}
\includegraphics[angle=0, width=1.0\linewidth]{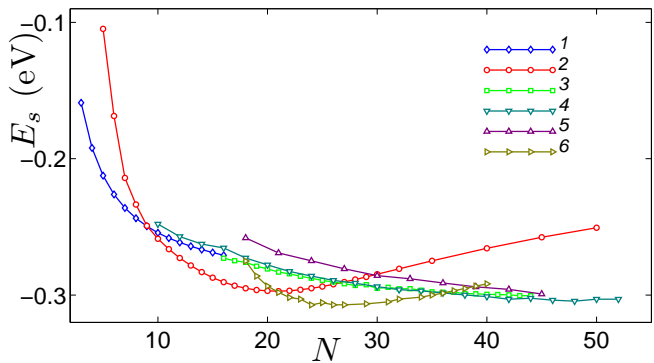}
\end{center}
\caption{\label{fg06}\protect
Dependence of the specific energy of the secondary structure of the homochiral ACN molecules $E_s$
on the number of molecules $N$ for an arc, a circle, one-beam, two-beam, three-beam spiral
and nested two circles (curves 1, 2, 3, 4, 5 and 6).
For the structure of two nested circles (curve 6), the dependence on the number of atoms
of the outer circle is shown.
}
\end{figure}

Hydrogen bond chains of $N\le 16$ ACN molecules of the same chirality form circular arcs
of the same radius. The step of such a chain (the distance between the oxygen atoms of neighboring
peptide groups) is $a=4.76$~\AA, the angle between neighboring links is $\varphi=162^\circ$,
the radius of curvature by oxygen atoms is $R=15.2$~\AA~ -- see Fig.~\ref{fg05}~(a).
The specific energy of the chain $E_s=E/N$ decreases monotonically with the growth of the number
of molecules $N$ -- see Fig.~\ref{fg06}.
When the number of links is $N>16$, the arcs cease to be stable; they either close
and form circular chains, or touch their ends and form flat spirals -- see Fig.~\ref{fg05} (b), (c).
\begin{figure}[tb]
\begin{center}
\includegraphics[angle=0, width=1.0\linewidth]{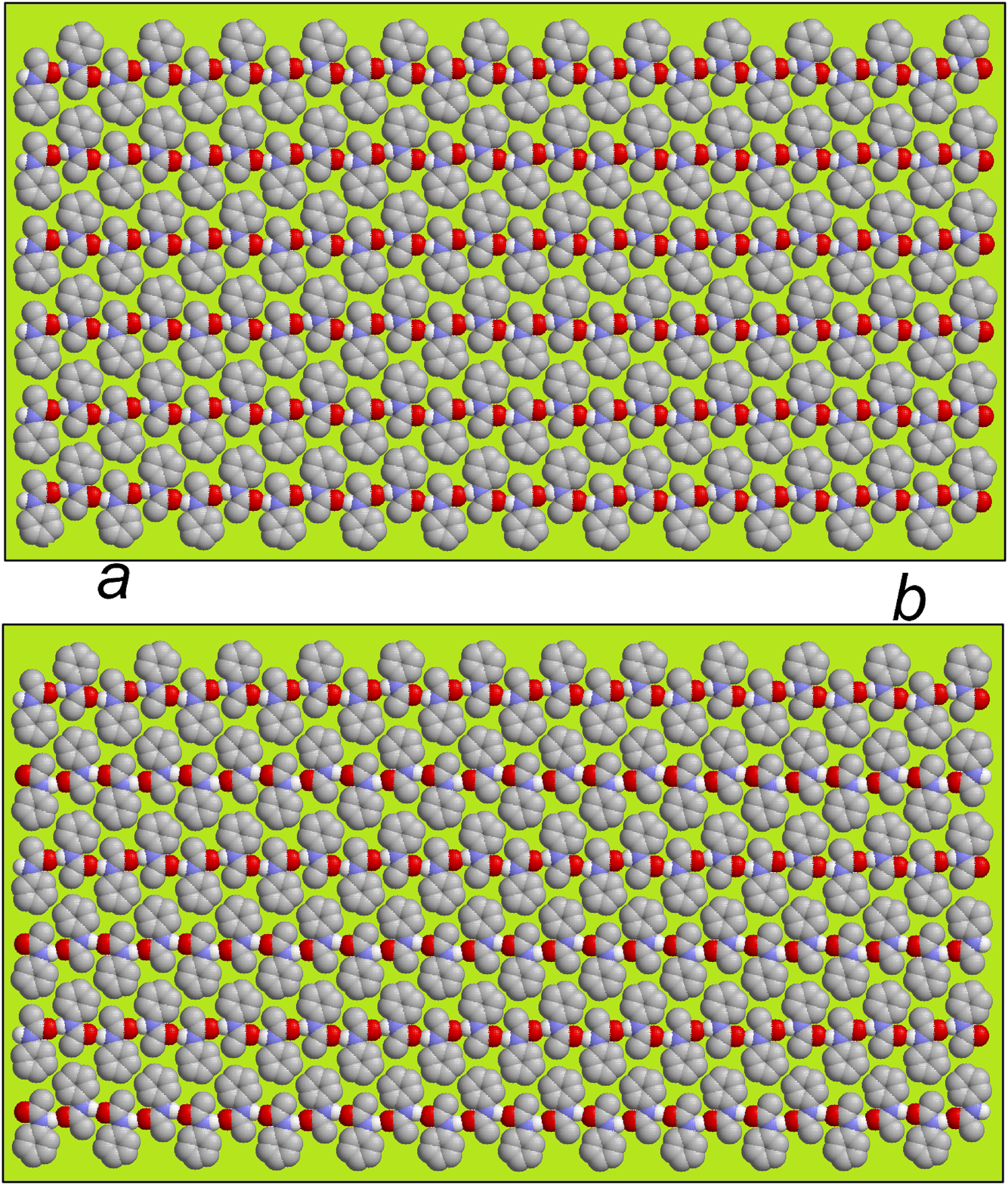}
\end{center}
\caption{\label{fg07}\protect
2D crystals of ACN molecules on a flat substrate:
(a) with parallel packing of linear chains of hydrogen bonds (periods $a_x=9.81$, $a_y=10.24$),
(b) with antiparallel packing ($a_x=9.79$, $a_y=20.43 $~\AA).
On the surface of a flat substrate, the chain of hydrogen bonds can be linear only
if the L and R isomers alternate.
A crystal is formed by 6 linear chains of 24 molecules (the total number of molecules is $N=6\times24$).
The flat substrate is shown in green.
}
\end{figure}

Stable cyclic chains can be formed from $N\ge 5$ molecules of the same chirality.
The dependence of the specific energy of the cyclic chain $E_s$ on the number of its links $N$
is shown in Fig.~\ref{fg06}. As can be seen from the figure, the most energetically favorable
are cyclic chains of $N=20$, 21, 22 links.
Such chains form ring structures with inner $R_1=27.3$, 28.8, 30.5 and outer radii
$R_2=40.4$, 41.9, 43.6~\AA.

The dependence $E_s(N)$ for a one-beam spiral actually continues the dependence for an arc
(see Fig.~\ref{fg06}, curves 1 and 3).
Two-beam and three-beam spirals are bound states of arc structures.
The specific energy of helical structures decreases monotonically with an increase in the number
of molecules.
For $N>27$, helical structures are more energy efficient than ring structures
(this is due to their denser structure).

The most energy-efficient are nested structures of two circles with the number of atoms
$N=14+26$, 15+27, 17+29 (the first number in the sum corresponds the number of atoms in the
inner circle, the second -- in the outer circle) -- see Fig.~\ref{fg06}, curve 6.

If the chain of hydrogen bonds consists of a random sequence of isomers,
then it will look like an irregular broken line.
The chain has the shape of a straight zigzag only if there is a strict alternation
of L and R isomers. In this case, the zigzag step (the distance between the oxygen atoms
of neighboring molecules) is $a=4.85$~\AA, the zigzag angle is $\varphi=170^\circ$.
Such chains on a flat substrate surface can form two-dimensional regular lattices (2D crystals)
with parallel and antiparallel packing of neighboring chains -- see Fig.~\ref{fg07}.
With parallel packing, the crystal periods are $a_x=9.81$\AA, $a_y=10.24$\AA,
the specific energy is $E_s=-0.325$~eV.
With antiparallel packing, periods are $a_x=9.79$\AA, $a_y=20.43$\AA, energy is $E_s=-0.322$~eV.
\begin{figure}[tb]
\begin{center}
\includegraphics[angle=0, width=1.0\linewidth]{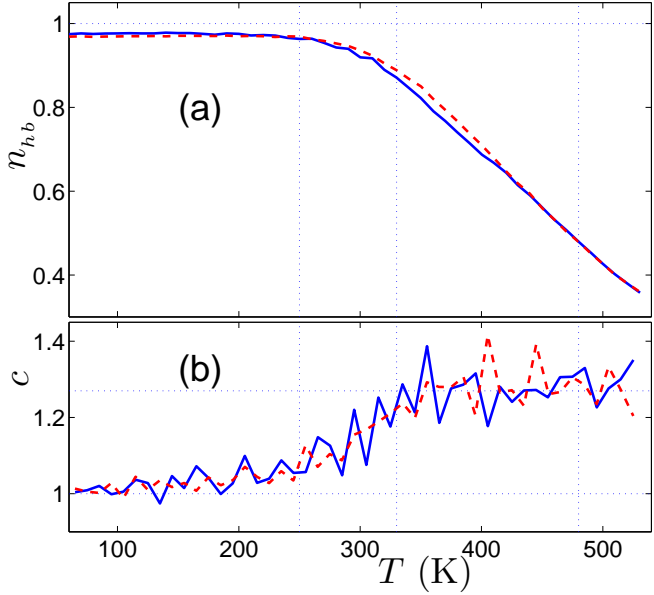}
\end{center}
\caption{\label{fg08}\protect
Dependence of (a) the normalized number of hydrogen bonds $n_{hb}$ and
(b) the dimensionless heat capacity $c$ on temperature $T$ for a system
of $N=1024$ ACN molecules located on a flat substrate with the periodic
square computational cell of size $33\times33$~nm$^2$.
Solid (blue) lines show dependencies for a system of homochiral molecules,
dotted (red) lines -- dependencies for a racemic mixture of molecules.
Vertical dotted straight lines correspond to temperatures $T=250$, 330, and 480~K.
}
\end{figure}

\section{Self-assembly of molecular structures}

Let us simulate the self-organization of the molecular structures of ACN molecules
on the flat surface of the h-BN crystal.
To do this, we take a square periodic cell of size $33\times33$~nm$^2$
on the surface of the substrate and randomly place $N=1024$ ACN molecules into it.
Then we immerse this molecular system in a Langevin thermostat of temperature $T$
and numerically simulate the dynamics of the system during the time $t=10$~ns.
To do this, we numerically integrate the system of Langevin equations
\begin{equation}
{\bf M}\ddot{\bf X}_n=-\frac{\partial}{\partial{\bf X}_n} H-\Gamma{\bf M}\dot{\bf X}_n-\Xi_n,~~n=1,...,N,
\label{f10}
\end{equation}
where $\Gamma=1/t_r$ is the friction coefficient, $\Xi_n=\{\xi_{n,i,k}\}_{i=1,k=1}^{11,~3}$
is 33-dimensional vector of normally distributed random Langevin forces with the
following correlations:
$$
\langle\xi_{n_1,i,k}(t_1)\xi_{n_2,j,l}(t_2)\rangle
=2M_{i}k_BT\Gamma\delta_{n_1n_2}\delta_{ij}\delta_{kl}\delta(t_1-t_2).
$$
Here $M_{i}$ is mass of $i$-th atom of ACN molecule,
$k_B$ is Boltzmann constant, $T$ is temperature
of the Langevin thermostat (temperature of the substrate),
numbers $n_1,n_2=1,...,N$, $i,j=1,...,11$, $k,l=1,2,3$.

The parameter $t_r$ characterizes the intensity of energy exchange between the molecular system
and the thermostat.
Simulation of the dynamics of ACN molecules on an h-BN sheet, taking into account the mobility
of the sheet atoms, makes it possible to estimate the relaxation time $t_r\sim 100$~ps.
For the convenience of numerical integration, we will use a smaller value $t_r=10$~ps.
This makes it possible to significantly reduce the time of numerical integration,
which is sufficient to obtain reliable average values.
After the dynamics of the molecular system reaches the steady state, we will find the time averages
of the system energy $\bar{E}(T)$ and the number of hydrogen bonds $\bar{N}_{hb}(T)$.
We assume that two ACN molecules form a hydrogen bond if their interaction energy
is greater than half of the hydrogen bond energy: $U({\bf X}_1,{\bf X}_2)<-E_{hb}/2=-0.16$~eV.

The state of the system can be conveniently characterized by its dimensionless heat capacity
\begin{equation}
c=\frac{1}{33Nk_B}\frac{d\bar{E}(T)}{dT},
\label{f11}
\end{equation}
and the normalized number of hydrogen bonds $n_{hb}=\bar{N}_{hb}(T)/N$.
The dependence of these quantities on temperature is shown in Fig.~\ref{fg08}.
\begin{figure}[tb]
\begin{center}
\includegraphics[angle=0, width=1.0\linewidth]{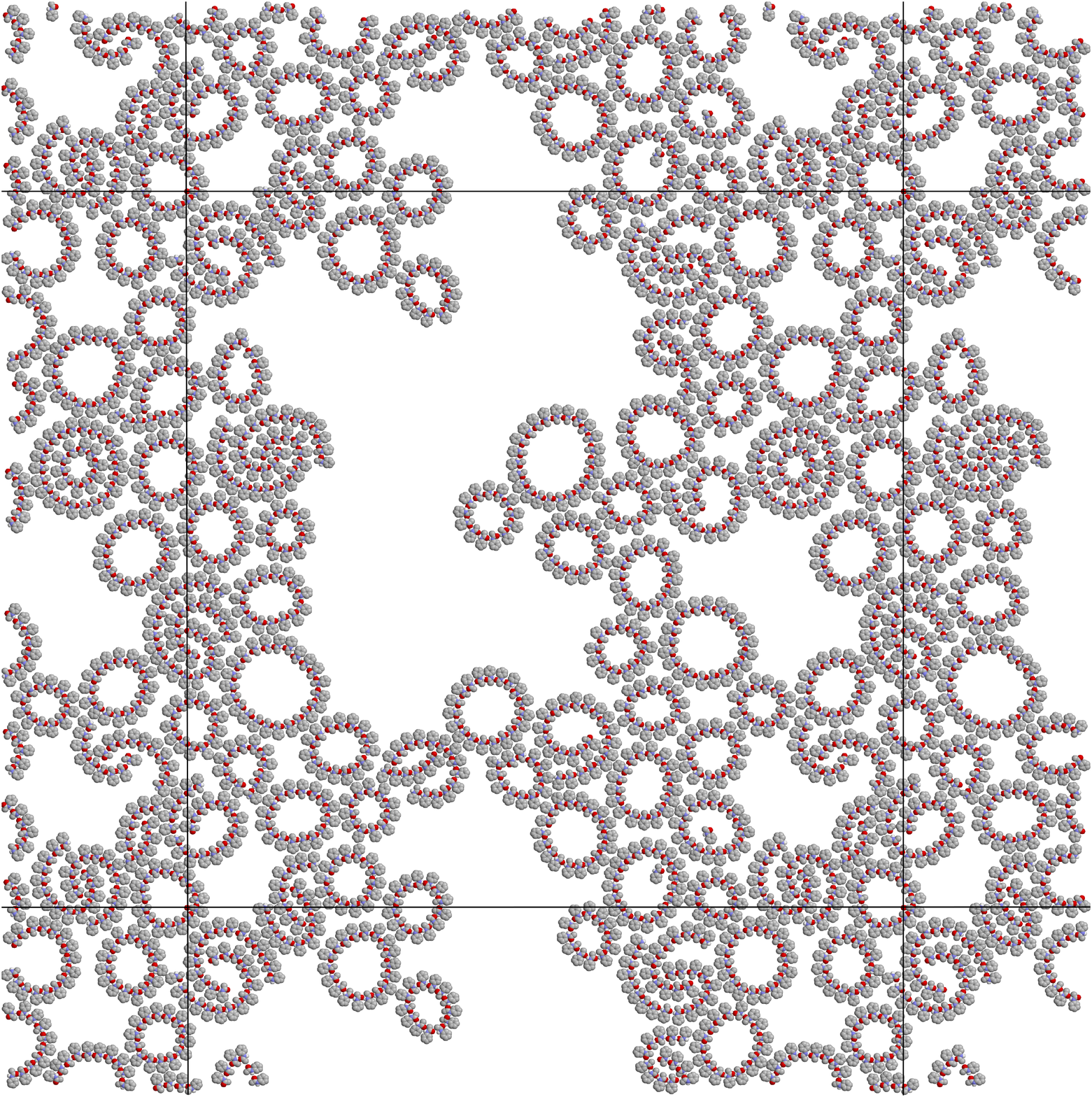}
\end{center}
\caption{\label{fg09}\protect
Structure of $N=1024$ homochiral ACN molecules appearing on the flat surface of the substrate
at $T=240$K. The straight lines show the boundaries of the periodic calculation cell of size
$33\times 33$~nm$^2$. The substrate surface is not shown.
}
\end{figure}
\begin{figure}[tb]
\begin{center}
\includegraphics[angle=0, width=1.0\linewidth]{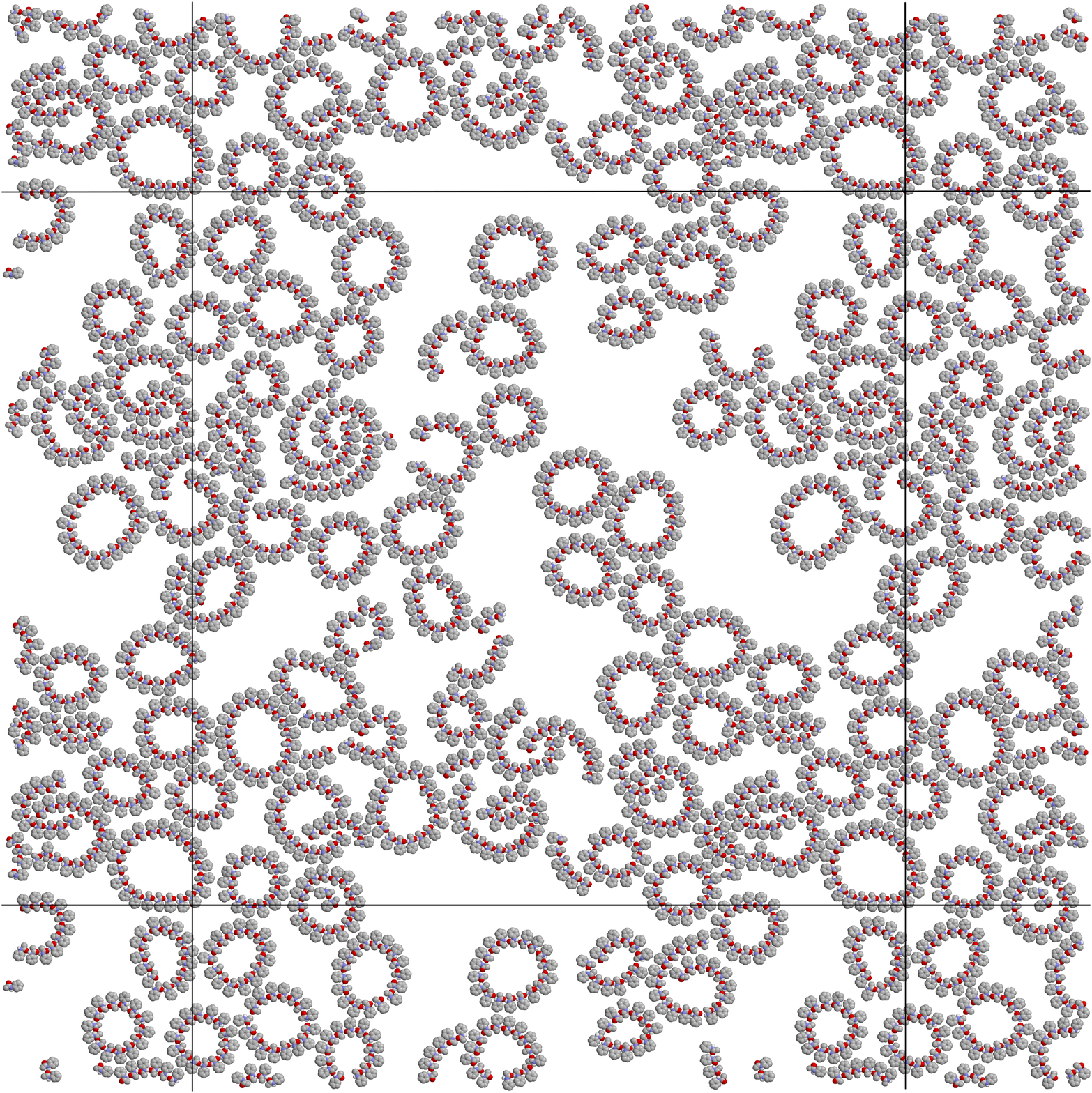}
\end{center}
\caption{\label{fg10}\protect
Structure of $N=1024$ homochiral ACN molecules appearing on the flat surface of the substrate
at $T=300$K. The straight lines show the boundaries of the periodic calculation cell of size
$33\times 33$~nm$^2$.
}
\end{figure}
\begin{figure}[tb]
\begin{center}
\includegraphics[angle=0, width=1.0\linewidth]{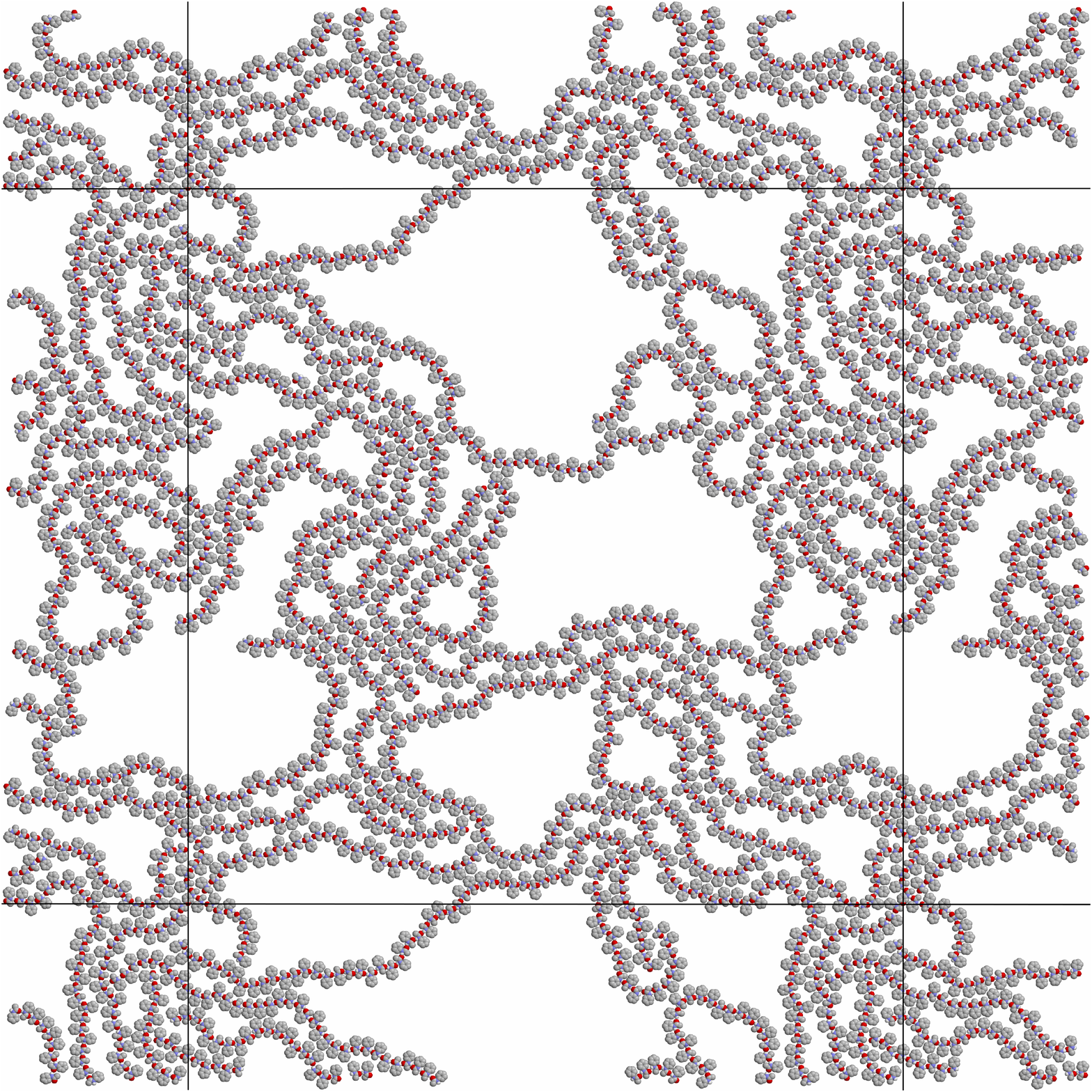}
\end{center}
\caption{\label{fg11}\protect
Structure formed on the flat surface of the substrate at $T=240$K from a racemic mixture of
$N=1024$ ACN molecules. The straight lines show the boundaries of the periodic calculation cell of size
$33\times 33$~nm$^2$.
}
\end{figure}

Numerical simulation shows the existence of three characteristic temperature values $T_1<T_2<T_3$.
At $T<T_1=250$K the molecules on a flat substrate form a stable system of chains of hydrogen bonds.
Here, almost every molecule participates in the formation of one hydrogen bond
(number $n_{hb}\approx 1$). The dimensionless heat capacity of the system is $c=1$.
At $T_1<T<T_2=330$K, a slight decrease in the number of $n_{hb}$ bonds and a monotonous increase
in heat capacity begin to occur -- the process of melting of hydrogen bond chains begins.
At a temperature $T>T_2$, the number of hydrogen bonds decreases in proportion to the increase
in temperature, and the heat capacity reaches a constant value $c\approx 1.23$.
Here we have a melt of short chains of hydrogen bonds (the average length of the chains decreases
proportionally to the increase in temperature). At $T>T_3=480$K, individual molecules can already
be detached from the substrate -- desorption of molecules begins.

The structure of the resulting system of hydrogen bond chains depends on the isomeric composition
of the molecules. If all molecules are homochiral (only one isomer is present),
then secondary structures form on the surface of the substrate.
These structures are circular and spiral hydrogen bond chains -- see Figs.~\ref{fg09} and \ref{fg10}.
The substrate surface become optically active, the left isomers form chains with a right twist,
and the right ones with a left twist.
As can be seen from Fig.~\ref{fg09} for $T=240$K on a flat substrate
all possible circular secondary structures (arcs, circles, spirals) are formed (Fig.~\ref{fg05}).
An increase in temperature leads, first of all, to the destruction of spiral structures.
As a result, the number of cyclic chains of average radius increases since they
are the most stable -- see Fig.~\ref{fg10}.

If the isomers of molecules are taken randomly, we get their racemic mixture.
In this case, no secondary structures are formed.
There are only curvilinear chains of random shape -- see Fig.~\ref{fg11}.
\begin{figure}[tb]
\begin{center}
\includegraphics[angle=0, width=1.0\linewidth]{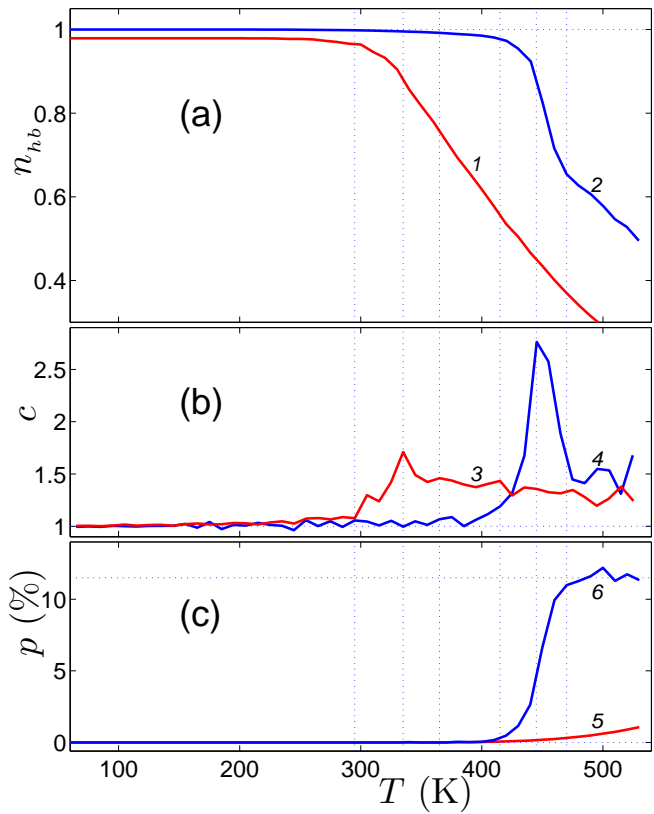}
\end{center}
\caption{\label{fg12}\protect
Dependence of (a) the normalized number of hydrogen bonds $n_{hb}$,
(b) the dimensionless heat capacity $c$ and (c) the fraction of molecules
that left the first layer $p$ on the temperature $T$ for a 2D crystal of $N=1056$ ACN
molecules located on a flat substrate with periodic computational cell
of size $46.6\times45$~nm$^2$ (curves 1, 3, 5; 25\% substrate coverage)
and $23.3\times22.5$~nm$^2$ (curves 2, 4 , 6; 100\% substrate coverage).
Vertical dotted straight lines show temperature values $T=295$, 335, 365, 415, 445 and 470~K.
}
\end{figure}
\begin{figure}[tb]
\begin{center}
\includegraphics[angle=0, width=1.0\linewidth]{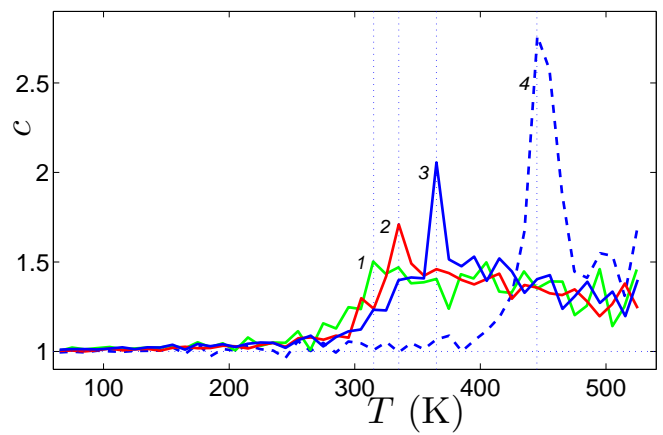}
\end{center}
\caption{\label{fg13}\protect
Dependence of the dimensionless heat capacity $c$ on temperature $T$
for a 2D square crystallite of $N=480$, 1056, 2376 of the ACN
molecules located on a flat substrate with periodic computational cell
of the size  $31.2\times 30.68$, $46.6\times45$, $70.2\times67.5$~nm$^2$
(curves 1, 3, 5; 25\% substrate coverage)
and for infinite  2D crystal (curve 4).
Vertical dotted straight lines mark the temperatures
$T=315$, 335, 365, and 445~K.
}
\end{figure}

\section{Melting of 2D crystals}

To simulate the dynamics of 2D crystals of ACN molecules on a flat h-BN substrate,
consider a crystal formed by 22 linear chains of hydrogen bonds of 48 molecules
(total number of molecules $N=22\times 48=1056$).
A crystal with parallel chain packing has dimensions of $23.3\times 22.5$~nm$^2$.
We place the crystal in the center of the calculated periodic cell size $23.5\times 22.6$~nm$^2$.
In this case, the chains of hydrogen bonds located parallel to the $x$ axis are closed,
and the first chain begins to come into contact with the last one -- the 2D crystal form
a dense packing on the substrate that has no edges and has normalized number of hydrogen bonds
$n_{hb}=1$ (the number of hydrogen bonds is equal to the number of molecules).
To simulate a rectangular crystal with edges (square crystallite), we take a periodic
computational cell of size $47\times 45.2$~nm$^2$.
In this case, the 2D crystal (crystallite) covers only 25\% substrate surface,
the chains of hydrogen bonds are not closed,
the normalized number of hydrogen bonds is $n_{hb}=(N-22)/N=0.979$.

Then we numerically integrate the system of equations of motion (\ref{f10})
with the initial condition corresponding to the stationary state of the crystal.
At different thermostat temperatures, we find the average values of the system energy $\bar{E}(T)$,
the normalized number of hydrogen bonds $n_{hb}$, and the fraction of molecules that left
the substrate from the first layer $p$ (the fraction of molecules located
on the substrate at distance $z>5$~\AA).
Next, using the formula (\ref{f11}), we find the temperature dependence
of the heat capacity of the molecular system $c$.

The dependence of $n_{hb}$, $c$, and $p$ on the thermostat temperature
(on the substrate temperature) $T$ is shown in Fig.~~\ref{fg12}.
As can be seen from the figure, at 25\% coverage of the substrate,
the square crystallite begins to melt at a temperature of $T_1=295$K.
At $T<T_1$, the crystal structure is preserved, the number of bonds, heat capacity,
and density of the first layer of the substrate coating practically do not change
with increasing temperature: $n_{hb}(T)\equiv n_{hb}(0)$, $c(T)\equiv 1$, $p(T)\equiv 0$.
Crystallite melting occurs in the temperature range $T_1<T<T_2$ where the upper temperature is $T_2=365$~K.
Here, a slight decrease in the number of bonds and an increase in heat capacity begin to occur.
The heat capacity reaches its maximum value at the temperature $T_m=335$~K.
In the temperature interval $[T_1,T_2]$, an increasing destruction of the edges
of the initial crystallite occurs. The ends of the chains peel off from the central part
of the crystallite, then break off and go to the free part of the substrate.
As a result of this "continuous"\ melting, the crystallite transforms into a melt consisting
of short chains of hydrogen bonds, uniformly covering the entire substrate.

For $T>T_2$, the number of hydrogen bonds decreases in proportion to the increase in temperature,
while the heat capacity remains almost constant: $c\approx 1.4$.
It can be concluded that a complete transition of the molecular system from a low-energy
and low-entropy crystalline state to a high-energy and high-entropy liquid state has taken place.
As a result of this transition, the substrate is uniformly covered with a "solution"\
of short chains of hydrogen bonds. The chain lengths decreases monotonically with increasing temperature
(on the Fig.~\ref{fg12}~(a)  this manifests itself in a monotonous decrease in the number of bonds).
All molecules remain directly adjacent to the substrate ($p=0$),
insignificant desorption is observed only at $T>480$~K -- see Fig.~\ref{fg12}~(c).

With 100\% coverage of the substrate, due to the absence of edges in the crystal,
its melting occurs at higher temperatures and happens according to a different scenario.
Here melting also occurs "continuously"\ in the temperature interval $[T_1,T_2]$, where $T_1=415$, $T_2=470$~K.
The peak of the heat capacity at $T_m=445$~K becomes more pronounced.
Melting occurs due to the expulsion of some molecules from the first layer to the second,
which manifests itself in a monotonous increase in the fraction of displaced molecules $p$
at $T_1<T<T_2$ -- see Fig.~\ref{fg12}~(c).
As a result, the density of the first layer during melting decreases by 11\%,
and after melting a dense melt of molecules is formed on the substrate,
in which 11\% of the molecules are located on the second layer from the substrate.

Conventionally, the temperature value $T_m$, at which the heat capacity has reached its maximum value,
can be considered as the melting temperature of a 2D crystal.
However melting occurs not discretely, but continuously in the temperature interval $[T_1,T_2]$.
The value $T_m$ is in the center of this interval.

To study the dependence of the melting temperature on the crystallite size,
we analyze numerically the melting of 2D square crystallite of $N=480$ and 2376 ACN molecules placed on a flat substrate with periodic computational cell of size $31.2\times 30.68$ and $70.2\times67.5$~nm$^2$. Our results show that melting occurs continuously for all crystallite sizes.
When the crystallite size is increased, the melting interval  $[T_1,T_2]$ shifts to the right and, in the limit, coincides with the melting temperature interval of a 2D crystal  with 100\% coverage of the substrate, as shown in Fig.~\ref{fg13}.

Let us note that the continuum melting scenario also takes place for 2D n-alkane crystals lying on a flat surface of graphite \cite{Zhang2021}.
This allows us to conclude that the quasi-continuous melting scenario is a characteristic feature of 2D systems of molecules adsorbed by a flat surface.

\section{Conclusion}

Our numerical simulations of the dynamics of the system of acetanilide molecules have revealed that the structures achiral in three-dimensional space become chiral when being placed on a flat substrate: Depending on the side it touches the substrate, the molecule has two isomers L and R. The homochirality of the molecules leads to the appearance of stable secondary structures stabilized by hydrogen bonds on a flat substrate in the form of arc, cyclic, and helical chains of hydrogen bonds and their complexes.

Hydrogen-bond chains of $N\le 16$ molecules of the same chirality form circular arcs of the same radius. When the number of molecules in the chain is $N>16$, the arcs becomes unstable, they either collapse into circles, or touch their ends and form flat spirals. In addition to single-beam spirals, two- and three-beam spirals may exist being bound states of arc chains.

Stable cyclic chains can be formed from $N\ge 5$ molecules of the same chirality.
The most energetically favorable are cyclic chains of 20, 21 and 22 links.
The structures of two circles with the number of atoms $N=14+26$, $15+27$, $17+29$
(where the first number of the sum is the number of atoms in the inner circle,
and the second number is the number of atoms in the outer circle)  are more energy efficient.

If the chain of hydrogen bonds consists of a random sequence of isomers,
it look like an irregular broken line. A chain can take the form of a rectilinear zigzag only if there is a strict alternation of the L and R isomers. Such chains can form regular two-dimensional crystals with parallel and antiparallel packing of the adjacent chains.

Simulation of the dynamics of a system of molecules shows that the homochirality of molecules(the presence of only one isomer) leads to the appearance of stable secondary structures on the surface of the substrate, i.e. to the appearance of chains of hydrogen bonds in the form of arcs, circles and spirals. As a result, the substrate surface becomes optically active, the left isomers form chains with a right twist, and the right ones form chain with a left twist. At temperature $T\le 240$~K, all possible secondary structures are formed on the substrate. An increase in temperature leads, first of all, to the disintegration of spiral structures.
As a result, the number of more stable circular chains increases.

If the molecules on the substrate form a racemic mixture, no regular secondary structures are formed, so only curvilinear chains of hydrogen bonds of random shape can appear. Thus, our results demonstrate the importance of homochirality (chiral purity) of biomolecules for the formation of stable secondary molecular structures.

Numerical simulations of the dynamics of a 2D crystal of the ACN molecules shows that the scenario of crystal melting depends on the density of its coverage of the substrate.

At 25\% coverage of the substrate the melting occurs "continuously"\ in the temperature interval $[T_1,T_2]$ from the edges of the initial crystallite (for crystallite of 1056 ACN molecules temperatures $T_1=295$, $T_2=365$~K).
The ends of the chains peel off from the central part of the crystallite, then break off and go to the free part of the substrate. As a result, the crystallite transforms into a melt consisting of short chains of hydrogen bonds, uniformly covering the entire substrate. When the size of the crystallite is increased, the melting interval  $[T_1,T_2]$ shifts to the right and, in the limit, will coincides with the melting temperature interval of infinite crystal
with 100\% coverage of the substrate.

For 100\% coverage of the substrate with no crystal edges, the melting occurs at higher temperatures $415\div 470$~K by a shift of some molecules from the first to the second layer in the substrate. As a result, the density of the first layer during melting decreases by 11\%, and after melting, 11\% of the molecules move to the second layer formed on the substrate.

A continual melting scenario (the presence of a melting temperature interval)has also been found in 2D n-alkane crystals lying on a flat surface of graphite \cite{Zhang2021}. All this leads us to conclusion that the quasi-continuous melting scenario is a characteristic feature of 2D systems of molecules adsorbed by a flat surface.

\begin{center}
{\bf ACKNOWLEDGMENTS}
\end{center}
AVS acknowledges the use of the computational facilities provided by the Interdepartmental Supercomputer Center of the Russian Academy of Science. YSK acknowledges a support from the Australian Research Council (grant DP210101292)
and the Strategic Fund of the Australian National University.

\end{document}